\documentclass[aps,amsfonts,amsmath,superscriptaddress,twocolumn,showpacs,floatfix,nofootinbib,showkeys]{revtex4-1}

\usepackage{marginnote}

\usepackage{slashed}
\usepackage{color, verbatim}
\usepackage{latexsym}
\usepackage{epsfig}
\usepackage{amsmath}
\usepackage{caption}

\usepackage{stackrel}

\usepackage{amssymb}
\usepackage{graphicx}
\usepackage{bm}

\usepackage{hyperref}
\usepackage{epstopdf}
\definecolor{myred}{rgb}{0.7, 0, 0}
\definecolor{myblue}{rgb}{0, 0, 0.7}
\definecolor{mygreen}{rgb}{0.04, 0.7, 0.5}
\hypersetup{colorlinks,citecolor=myred,linkcolor=myblue,urlcolor=myblue,linktocpage=true}

\makeatletter

\newcommand{\be}{\begin{equation}}
\newcommand{\ee}{\end{equation}}
\newcommand{\bea}{\begin{eqnarray}}
\newcommand{\eea}{\end{eqnarray}}

\begin{document}

\title{Weakly Inhomogeneous models for the Low-redshift Universe}

\author{P. Marcoccia }
\email{marcocciapaolo1991@gmail.com}
\affiliation{Department of Mathematics and Physics, University of Stavanger, NO-4036 Stavanger, Norway\\
             Dipartimento di Fisica Università "La Sapienza", Piazza A.Moro 5, 00185 Roma, Italy\\
             Tel.: +39 3284084152\\}

\author{G. Montani}
\email{giovanni.montani@enea.it}
\affiliation{Dipartimento di Fisica Università "La Sapienza", Piazza A.Moro 5, 00185 Roma, Italy\\
       ENEA, Fusion and Nuclear Safety Department C.R. Frascati, Via E. Fermi 45 (00044) Frascati (RM), Italy\\
       Tel.: +39  06 49914536\\}

\date{\today}

\begin{abstract} 
\noindent We analyze two different algorithms for 
constructing weakly inhomogeneous models 
for the low-redshift Universe, 
in order to provide a tool for 
testing the geodesic dynamics, within
the sphere of validity for the 
Universe acceleration. 
We first implement the so-called 
quasi-isotropic solution in the 
late Universe, when a pure dark 
energy equation of state for the 
cosmological perfect fluid is considered. 
We demonstrate, that a solution exists 
only if the physical scale of the inhomogeneities is larger then the Hubble
scale of the microphysics,
which implies that inhomogeneities could not
be observed at present-stage time.
Then, we analyze a weakly deformed isotropic Universe toward a spherically
symmetric model, thought as the 
natural metric framework of the 
$\Lambda CDM$ model.
We find that inhomogeneities will arise in the previous context as small perturbations 
of the order of few point percent over the background FRW universe.
The obtained picture offers a useful 
scenario to investigate the influence 
of the inhomogeneity spectrum 
(left free in the obtained solution), 
on the propagation of photons or gravitational waves at low 
redshift values, and in line of principle may be used to account for several present-stage cosmological problems, such as the Hubble Tension.\\ 
\end{abstract}



\maketitle



\section{Introduction}
\label{intro}
\noindent The Standard Cosmological model \cite{Montani:2011zz}\cite{Kolb:1990vq}, is based on the homogeneous and isotropic \emph{Robertson-Walker} metric, 
and bases its reliability on the high isotropy of the \emph{Cosmic Microwave Background Radiation} \cite{Komatsu:2014ioa}\cite{Ade:2015xua}.

Actually, the estimates from the galaxy surveys of the spatial scale at which the present Universe reaches homogeneity, 
provides a value of about $60 Mpc/h$ \cite{Ntelis:2016utg}, where $h \approx 0.7$,  which implies that, at lower scales, 
significant deviations from the Robertson-Walker geometry may be observed, at least as higher order corrections. 
The presence of such small scale deviations could influence the information we get from extragalactic sources and, 
in general, they affect the particles' path from distant regions up to our detectors \cite{Buchert:2015iva}.

In this work, we investigate the spatial metric which admit inhomogeneous corrections to the flat isotropic model 
(the present contribution of the spatial curvature is clearly negligible with respect to the matter terms).
As a first step, we analyze the so-called quasi-isotropic solution \cite{Lifshitz:1963ps} (see also \cite{Montani:1999ez}\cite{Imponente:2003ds}), 
but implemented in the low redshift Universe.
This solution is a general one of the \emph{Einstein equations} \cite{1915SPAW.......778E}, which admits inhomogeneous perturbations to the background metric in the form of a first order term factorized as a time dependant part times a function of the pure $3$ spatial coordinates, the downside of this freedom is a much more complicated form for the resulting field equations.
In particular, we study this solution in the presence of a perfect fluid, having a \emph{dark energy} \cite{Frieman:2008sn} equation of state,
i.e. $P = w\rho$ ($P$ and $\rho$ being the pressure and energy density of the fluid respectively),
where $-1 < w < -1/3$.
The interesting feature of the obtained solution, relies on the fact that the considered inhomogeneities, included in the model as small corrections, 
must correspond, in order for the solution be consistent, to physical scales much greater than the \emph{Hubble (microphysical) scale}\footnote{Given a point in space $A$, the Hubble length may be considered as the radial distance of the points 
that, due to the Hubble expansion, recedes from $A$ with a speed equals to $c$, where $c$ is the speed of light.}. 
This results in inhomogeneous corrections being of pure curvature nature and, consequently, they can not affect the physical processes taking place in the Hubble sphere.

Then, we consider the case of a \emph{Lemaitre-Tolman-Bondi} spacetime \cite{Montani:2011zz}\cite{Peebles:1994xt},
which is again an inhomogeneous solution of the \emph{Einstein equations} but, unlike the \emph{Lifshitz-Khalatnikov} quasi-isotropic solution \cite{Lifshitz:1963ps}, may be isotropic by observing the universe from a preferred point that is singled out during the construction of the model.
The main reason why the \emph{LTB} may appear isotropic but inhomogeneous lies in the fact that in this approach the inhomogeneities of the model are described as function of the sole radial coordinate $r$, this will indeed lead to a loss of generality that will be trade-off for a much simpler form of the field equations.
We use such a solution to describe a spherically symmetric Universe in the presence of a matter source, 
and a non-zero \emph{cosmological constant},  
an appropriate scenario to account for the so-called $\Lambda CDM$ model \cite{Riess:1998cb}. 
Clearly, in order to describe the local behavior of the actual Universe, as well as to make consistent the obtained model with the previously described quasi-isotropic solution, we consider the inhomogeneous perturbations again 
as a first order modification of the flat Robertson-Walker geometry.
We demonstrate the existence of a consistent solution of the linearized \emph{Einstein equations}, 
which does not fix the radial dependence of the inhomogeneities, but only their time scaling.
Moreover, we specialize the obtained solution to the case of the $\Lambda CDM$ model, by tuning the values of the parameters in order to obtain, 
that the matter be the $30\%$ and the constant energy density the $70\%$ of the Universe critical parameter respectively.
The obtained time profile for the perturbations, together with the arbitrariness of their specific spatial morphology, 
offer an interesting arena to study the effects on the propagation of photons or gravitational waves, due to the local deviations of the actual Universe from homogenity.

Furthermore, the results of our analysis suggest an intriguing issue: 
while the inhomogeneities allowed by a $\Lambda CDM$ model are physically observable, living in principle, in the present Hubble sphere, 
the dark energy dominated Universe appears incompatible with the physical scale of inhomogeneity, 
the microphysics processes remain essentially concerned by the homogeneity restriction.
The different behavior of the two considered equations of state (we recall that the cosmological constant is associated to the relation $P = -\rho$), 
could become a qualitative discrimination property when incoming missions, like \emph{Euclid} \cite{Amendola:2016saw} will be able to test the large scale properties of the Universe, 
detecting details of the matter distribution across the cosmological space.

This paper is structured as follows:
in section \ref{sec:2.}, there will be a review of the results obtained in literature for the \emph{Lifshitz-Khalatnikov} quasi isotropic solution, in particicular, the case of a pure radiation high-redshift universe will be described, so that in section \ref{sec:3.} the same procedure will be applied in the new context of a low-redshift dark energy universe.
Following the same steps, in section \ref{sec:4.} we illustrate the basic theory of a \emph{Lemaitre-Tolman-Bondi} model for a spherically symmetric universe in the generic case.
This result will then be used in section \ref{sec:5.} to extend the previous solution to the case of a low redshift universe, filled with both a matter and a cosmological constant perfect fluid.
In particular, in subsection \ref{ssec:1.} the model obtained in section \ref{sec:5.} will be fitted with the actual observational data, in order to select the parameters that best describe the behaviour of our $\Lambda CDM$ universe, while in subsection \ref{ssec:2.} we will briefly discuss how the inhomogeneous perturbations affect the estimated value of the \emph{Hubble Constant}.
Lastly, the article will be closed with concluding remarks, that are reported in section \ref{sec:8.}.

\section{The \emph{Lifshitz-Khalatnikov} quasi-isotropic solution for \emph{high-redshift} radiation universes}
\label{sec:2.}

\noindent The \emph{Lifshitz-Khalatnikov} quasi-isotropic solution \cite{Lifshitz:1963ps}, is a generalization of the \emph{FRW} cosmology \cite{Friedmann}, in which a certain degree of inhomogeneity, and so anisotropy, is introduced. The inhomogeneity of space is reflected to the presence of three phisically arbitrary functions of the coordinates in the metric of the system.

In an isotropic solution, isotropy and homogeneity implies the vanishing of the off-diagonal metric components $g_{0 \alpha}$, 
while if isotropy and homogeneity assumption are dropped, it is always possible to move to a frame where the previous condition may be imposed. To do so, we must define a \emph{Synchronous reference frame} \cite{Landau:1982dva}, with the following choice for the metric tensor:
\begin{equation}
\begin{cases} g_{00} = 1 \\ g_{0 \alpha} = 0 \end{cases},
\label{eq:equation2.1}
\end{equation}
such that the metric reduces to the form:
\begin{equation}
ds^2 = dt^2 - h_{\alpha \beta}(x,t)dx^{\alpha}dx^{\beta},
\label{eq:equation2.2}
\end{equation}
where the term $h_{\alpha \beta}$, is called \emph{Trimetric}, and represent the pure spatial component of the metric.
In equation \eqref{eq:equation2.2}, as well as for the rest of the paper, we will work with \emph{natural units} \cite{Hsu:2011mf} by imposing $c=1$. 

The original \emph{Lifshitz-Khalatnikov} model was developed for a pure radiation universe,
and for the ultrarelativistic matter, the \emph{equation of state} reads as $P = \rho/3$, while the trimetric $h_{\alpha \beta}$ is linear in $t$ at the first order.

When searching for a quasi-isotropic extension of the \emph{Robertson-Walker} geometry, the metric should be expandable in integer powers of $t$, asymptotically as $t \rightarrow 0$,
following the Taylor-like expansion:
\begin{equation}
h_{\alpha \beta} = \sum^{\infty}_{n = 0} \left( \frac{\partial^{n} h_{\alpha \beta}}{\partial t^n} \vert_{t=0} \right) t^n_{0} \left( \frac{t}{t_{0}} \right)^n .
\label{eq:equation2.3}
\end{equation}
It might be observed, that $t_{0}$ is an arbitrary time that satisfies the condition $t << t_{0}$, while the existence of the cosmologic singularity implies the vanishing of the $0$ order term of the series.
For the following analysis of the model, only the first two terms of the series will be considered, so by defining:
\begin{equation}
a^{(n)}_{\alpha \beta} = \left( \frac{\partial^{n} h_{\alpha \beta}}{\partial t^n} \vert_{t=0} \right) t^n_{0} ,
\label{eq:equation2.4}
\end{equation}
it is possible to write:
\begin{equation}
h_{\alpha \beta} = a^{(1)}_{\alpha \beta} \frac{t}{t_{0}} + a^{(2)}_{\alpha \beta} \left( \frac{t}{t_{0}} \right)^2 + ...
\label{eq:equation2.5}
\end{equation}
After a suitable rescaling, and introducing the adimensional time $\tilde{t} = t/t_{0}$, the trimetric gets the form:
\begin{equation}
\begin{cases} h_{\alpha \beta} = a_{\alpha \beta} \tilde{t} + b_{\alpha \beta} \tilde{t}^2 + ...\\ h^{\alpha \beta} = a_{\alpha \beta} \tilde{t}^{-1} - b_{\alpha \beta} + ... \end{cases},
h^{\alpha \delta} h_{\delta \beta} = \delta^{\alpha}_{\beta} + ...
\label{eq:equation2.6}
\end{equation}
the system moreover, assumes a much easier form by defining the \emph{auxiliary tensors} $K_{\alpha \beta}$ as:
\begin{equation}
\begin{cases} K_{\alpha \beta} = \partial_{\tilde{t}} h_{\alpha \beta} = a_{\alpha \beta} + 2 \tilde{t} b_{\alpha \beta} \\
K^{\alpha}_{\beta} = h^{\alpha \delta} K_{\delta \beta} = \tilde{t}^{-1} \delta^{\alpha}_{\beta} + b^{\alpha}_{\beta}\\
K = \partial_{\tilde{t}} ln(h) = 3 \tilde{t}^{-1} + b \end{cases}
\label{eq:equation2.7}
\end{equation}
where from the third equation, it is possible to get:
\begin{equation}
h = det(h_{\alpha \beta}) \sim \tilde{t}^3 (1 + \tilde{t} b) det(a_{\alpha \beta}) .
\label{eq:equation2.8}
\end{equation}
The last thing needed to write down the \emph{Einstein Equations}, is the \emph{Energy-Momentum Tensor} \cite{1915SPAW.......778E}, that in the case of ultrarelativistic matter gets the form:
\begin{equation}
T_{ij} = \frac{\rho(t)}{3} (4 u_{i} u_{j} - g_{ij}) ,
\label{eq:equation2.9}
\end{equation}
leading to the following Einstein equations:
\begin{equation}
\begin{cases} R^{0}_{0} : \frac{1}{2} \partial_{\tilde{t}} K^{\alpha}_{\alpha} + \frac{1}{4} K^{\alpha}_{\beta} K^{\beta}_{\alpha} = -k \frac{\rho(\tilde{t})}{3} (4 u_{0}^2 - 1) \\
R^{0}_{\alpha} : \frac{1}{2} \left( K^{\beta}_{\alpha ; \beta} - K_{; \alpha} \right) =  \frac{4}{3} k \rho(\tilde{t}) u_{\alpha} u^{0}\\
R^{\alpha}_{\beta} : \frac{1}{2 \sqrt h} \partial_{\tilde{t}} \left( \sqrt h K^{\alpha}_{\beta} \right) + ^{3}R^{\alpha}_{\beta} = - k \frac{\rho(\tilde{t})}{3} (4 u_{\beta} u^{\alpha} + \delta^{\alpha}_{\beta}) \end{cases} 
\label{eq:equation2.10}
\end{equation}

In the last equations, we introduced the \emph{Einstein Constant} $k = 8 \pi G$, as well as the term $^{3}R^{\alpha}_{\beta}$ that is the \emph{Tridimensional curvature Ricci's Tensor}, which analitically assume the form:
\begin{equation}
^{3} R^{\alpha}_{\beta} =  \Gamma^{\gamma}_{\alpha \beta , \gamma} - \Gamma^{\delta}_{\beta \delta , \alpha} + \Gamma^{\sigma}_{\alpha \beta} \Gamma^{\lambda}_{\sigma \lambda}
- \Gamma^{\nu}_{\alpha \mu} \Gamma^{\mu}_{\beta \nu} ,
\label{eq:equation2.11}
\end{equation}
recalling the $4$\emph{-velocity} relation: 
\begin{equation}
1 = u_{i} u^{i} \sim u_{0}^2 - \tilde{t}^{-1} a^{\alpha \beta} u_{\beta} u_{\alpha} ,
\label{eq:equation2.12}
\end{equation}
and assuming that the approximation $u_{0}^2 \sim 1$ is valid, the system \eqref{eq:equation2.10} may be solved up to the zeroth $O(1/ \tilde{t}^{2})$ and first-order $O(1/ \tilde{t})$.

The solutions obtained are:
\begin{equation}
\begin{cases}  k \rho(\tilde{t}) = \frac{3}{4 \tilde{t}^2} - \frac{b}{2 \tilde{t}}\\
u_{\alpha} = \frac{\tilde{t}^2}{2} \left( b_{; \alpha} - b^{\beta}_{\alpha ; \beta} \right) \end{cases} .
\label{eq:equation2.13}
\end{equation}
From the second equation, it might be observed that the assumption $u_{0}^2 \sim 1$ is valid at the first order, de facto the second term of \eqref{eq:equation2.12} is proportional to $t^{3}$, and for $ t \rightarrow 0$ is negligible respect to the first term.

The \emph{density contrast} moreover, may be defined as the ratio between the perturbed term and the zeroth order term of the density, giving:
\begin{equation}
\delta = - \frac{2}{3} b \tilde{t} \sim t .
\label{eq:equation2.14}
\end{equation}
This behavior implies that, as expected in the standard cosmological model, the zeroth-order term of the energy density diverges more rapidly than the
perturbation and the singularity is naturally approached with a vanishing density contrast.

Considering the last remaining equation of \eqref{eq:equation2.10}, $^{3}R^{\alpha}_{\beta}$ at leading order may be written as:
\begin{equation}
^{3}R^{\alpha}_{\beta} = A^{\alpha}_{\beta}/t ,
\label{eq:equation2.15}
\end{equation}
on which $A^{\alpha}_{\beta}$ are pure functions of spatial coordinates, constructed in terms of $a_{\alpha \beta}$.

The third equation of \eqref{eq:equation2.10} so, upon using relation \eqref{eq:equation2.15}, reduces to:
\begin{equation}
A^{\alpha}_{\beta} + \frac{3}{4} b^{\alpha}_{\beta} + \frac{5}{12} b \delta^{\alpha}_{\beta} = 0 ,
\label{eq:equation2.16}
\end{equation}
which admits the following trace:
\begin{equation}
b^{\alpha}_{\beta} = \frac{4}{3} A^{\alpha}_{\beta} + \frac{5}{18} A \delta^{\alpha}_{\beta} .
\label{eq:equation2.17}
\end{equation}
Finally, using the \emph{tridimensional Bianchi identity}, that reads as:
\begin{equation}
A^{\alpha}_{\beta ; \alpha} = 1/2 A_{;\alpha} 
\label{eq:equation2.18}
\end{equation}
eq. \eqref{eq:equation2.17} reduces to:
\begin{equation}
b^{\alpha}_{\beta ; \alpha} = \frac{7}{9} b_{,\beta} 
\label{eq:equation2.19}
\end{equation}
and consequently, the tri-velocity components defined by the second equation of \eqref{eq:equation2.13}, assume the form:
\begin{equation}
u_{\alpha} = \frac{t^2}{9} b_{,\alpha} .
\label{eq:equation2.20}
\end{equation}
It might be worth observing, that the metric \eqref{eq:equation2.5} allows an arbitrary spatial coordinate transformation, while the above solution contains only
$3$ arbitrary space functions arising from $a_{\alpha \beta}$.\\

\section{The \emph{Lifshitz-Khalatnikov} quasi-isotropic solution for \emph{low-redshift} Dark Energy universes}
\label{sec:3.}

\noindent In this section we will now show how the \emph{Lifshitz-Khalatnikov} quasi-isotropic solution, may be applied for the analysis of \emph{low-redshift}, dark energy \cite{Frieman:2008sn} universes.

Let's start by introduce the \emph{scale factors' ratio} $\eta(t) = a^2(t)/b^2(t)$ in the trimetric of the system, obtaining:
\begin{equation}
\label{eq:equation3.1}
\begin{aligned}
h_{\alpha \beta}(t,x) &=  a^2(t) \xi_{\alpha \beta}(x^{\gamma}) +b^2(t) \theta_{\alpha \beta}(x^{\gamma}) + .... \\
&= a^2(t) \left[ \xi_{\alpha \beta}(x^{\gamma}) + \eta(t) \theta_{\alpha \beta}(x^{\gamma}) + ... \right] ,
\end{aligned}
\end{equation}
where $\xi_{\alpha \beta}(x^{\gamma})$ is the \emph{Minkowskian} unperturbed metric, and $ \theta_{\alpha \beta}(x^{\gamma})$ are free functions of spatial coordinates, that represent the inhomogeneous perturbation of the metric.

To analyze the system at low redshift, it must be imposed that the ratio $t/t_{0} >> 1$, where this time $t_{0}$ is used to denote an arbitrary time smaller than the present one, while the contrast $\eta(t)$ must respect the limit:
\begin{equation}
\underset{t \rightarrow \infty}{lim} \quad \eta(t) = 0 .
\label{eq:equation3.2}
\end{equation}
The condition \eqref{eq:equation3.2}, imposes that the perturbation scale factor $b(t)$ is negligibile in respect to the main scale factor $a(t)$ at odiern stage time, so that the inhomogeneous perturbation, will result in a small correction over the background homogeneous and isotropic universe.

Let's now consider the energy-momentum tensor, the equation of state for a dark energy fluid is $P = w \rho$ with $w \in \left(-1,-1/3\right]$, hence from now on we will parametrize $w = -1/3 - \delta_{w}$ with $\delta_{w}  \in \left[0,2/3 \right)$.
The energy-momentum tensor for a pure dark energy fluid, assumes the form:
\begin{equation}
\begin{cases} 
T_{00} = \rho \left(\frac{2}{3} - \delta_{w} \right) u_{0} u_{0} + \rho \left(\frac{1}{3} + \delta_{w} \right) g_{00} \\
T_{0 \alpha} = \rho \left(\frac{2}{3} - \delta_{w} \right) u_{0} u_{\alpha}\\
T_{ \alpha \beta} = \rho \left(\frac{2}{3} - \delta_{w} \right) u_{\alpha} u_{\beta} + \rho \left(\frac{1}{3} + \delta_{w} \right) g_{\alpha \beta} \\
T = \rho (2 + 3 \delta_{w}) \end{cases} 
\label{eq:equation3.3}
\end{equation}
while the Einstein equations, that in a synchronous reference frame gets the form:
\begin{equation}
\begin{cases} 
R^{0}_{0} : -\frac{1}{2} K_{,0} - \frac{1}{4} K^{\delta}_{\sigma} K^{\sigma}_{\delta} =  k(T^{0}_{0} - \frac{1}{2} T)\\
R^{0}_{\alpha} : - \frac{1}{2} \left( K_{; \alpha} - K^{\gamma}_{\alpha ; \gamma} \right)= k T^{0}_{\alpha}\\
R^{\alpha}_{\beta} : ^{3}R^{\alpha}_{\beta} + \frac{1}{2 \sqrt h} \left( \sqrt h K^{\alpha}_{\beta} \right)_{;0} = k \left(T^{\alpha}_{\beta} - \frac{1}{2} 
\delta^{\alpha}_{\beta} T \right) \end{cases} 
\label{eq:equation3.4}
\end{equation}
will reduce to:
\begin{equation}
\begin{cases} 
-\frac{1}{2} K_{,0} - \frac{1}{4} K^{\delta}_{\sigma} K^{\sigma}_{\delta} = k \rho \left[ \left( \frac{2}{3} - \delta_{w} \right)u_{0}^2 - \left( \frac{2}{3} + \frac{\delta_{w}}{2} \right) \right]\\
-\frac{1}{2} \left( K_{; \alpha} - K^{\gamma}_{\alpha ; \gamma} \right)= k \rho \left(\frac{2}{3} - \delta_{w} \right) u^{0} u_{\alpha}\\
^{3}R^{\alpha}_{\beta} + \frac{1}{2 \sqrt h} \left( \sqrt h K^{\alpha}_{\beta} \right)_{;0} =  k \rho \left[ \left( \frac{2}{3} - \delta_{w} \right)u^2 + \left( \frac{2}{3} + \frac{\delta_{w}}{2} \right) \delta^{\alpha}_{\beta} \right] \end{cases} 
\label{eq:equation3.5}
\end{equation}
In the third equation of \eqref{eq:equation3.5}, the term $u^2 = 1/a^2(t)(\xi^{\alpha \beta} u_{\alpha} u_{\beta})$ appears, following the standard \emph{Lifshitz-Khalatnikov} approach anyway, from now on it will be imposed $u^2 \approx 0$, reflecting to $u_{0} = 1$.
The \emph{tri-velocity}, that already has a temporal dependance of at least $O(1/t^2)$, appears in eqs. \eqref{eq:equation3.5} at least multiplied by the zeroth order term of the density, that in a \emph{FRW} cosmology goes as $O(1/t^2)$.
The validity of the approximation however, will be verified later by calculating the analytical form of tri-velocity $u_{\alpha}$.

To integrate the system \eqref{eq:equation3.5}, we must define the form that the auxiliary tensors assume, in relation to our current choice for the metric \eqref{eq:equation3.1}.
Following the definitions given in \eqref{eq:equation2.7}, at the first order they reduce to:
\begin{equation}
K_{\alpha \beta} = h_{\alpha \beta , 0} = 2 \dot{a}a \left[ \xi_{\alpha \beta} + \eta(t) \theta_{\alpha \beta} \right] + \dot{\eta}(t) a^2 (t) \theta^{\alpha}_{\beta} ,
\label{eq:equation3.6a}
\end{equation}
\begin{equation}
K^{\alpha}_{\beta} = h^{\alpha \delta} K_{\delta \beta} \approx  2 \frac{\dot{a}}{a} \delta^{\alpha}_{\beta} + \dot{ \eta }(t) \theta^{\alpha}_{\beta} ,
\label{eq:equation3.6b}
\end{equation}
\begin{equation}
K \approx 6 \frac{\dot{a}}{a} + \dot{ \eta} (t) \theta = \partial_{t} ln(h) .
\label{eq:equation3.6c}
\end{equation}
Furthermore, taking into account eq. \eqref{eq:equation3.6c}, the determinant of the trimetric $h_{\alpha \beta}$ get the form:
\begin{equation}
h = j a^6 (t) e^{\eta(t) \theta} ,  
\label{eq:equation3.7}
\end{equation}
that for $t >> 0$, under the assumption \eqref{eq:equation3.2}, may be approximated as:
\begin{equation}
h \approx j a^6 (t) (1 + \eta (t) \theta) \quad \Rightarrow \quad \sqrt{h} = a^3 (t) \left( 1 + \frac{\eta (t) \theta}{2} \right) . 
\label{eq:equation3.8}
\end{equation}
Moreover, working in the regime of $t >>t_{0}$ , and assuming that $t_{0} >>0$, the \emph{Tridimensional curvature Ricci's Tensor} may be approximated as:
\begin{equation}
^3 R^{\alpha}_{\beta} \approx \frac{\eta(t)}{a^2(t)} \theta^{\mu}_{\alpha , \beta , \mu} .
\label{eq:equation3.9}
\end{equation}
By assuming that both $a(t)$ and $b(t)$ follow a \emph{power law} relation in function of time, from eqs. \eqref{eq:equation3.5},\eqref{eq:equation3.6a},\eqref{eq:equation3.6b} and \eqref{eq:equation3.6c}, it might be observed that the system has $O(1/t^2)$ zeroth order terms, while has $O(\eta(t)/t^2)$ perturbation terms.
Knowing that, in a flat dark energy universe, the scale factor $a(t)$ evolves as:
\begin{equation}
a(t) = t^{\frac{2}{3(1 + w)}} ,
\label{eq:equation3.10}
\end{equation}
the tridimensional curvature Ricci's tensor \eqref{eq:equation3.9}, may be negligible in the eqs. \eqref{eq:equation3.5} when the condition:
\begin{equation}
a(t) >> t \quad \Rightarrow \quad t^{\frac{2}{3(1 + w)}} >> t^{1} \quad \Rightarrow \quad w < - \frac{1}{3} ,
\label{eq:equation3.11}
\end{equation}
is verified.

However, a condition must be imposed also on the pure spatial part of \eqref{eq:equation3.9}, and it may be done by introducing the \emph{scale of the perturbation}, that may be derived  from the equation:
\begin{equation}
\frac{\eta(t)}{a^2(t)} \theta^{\mu}_{\alpha , \beta , \mu} << \frac{\eta(t)}{t^2} .
\label{eq:equation3.extra}
\end{equation}
In fact, considering that $ \theta^{\mu}_{\alpha , \beta , \mu} \approx \theta^{\mu}_{\alpha}/ \lambda^2$ and $L_{H} \sim t$, where $L_{H}$ stands for the \emph{Hubble Length}, eq. \eqref{eq:equation3.extra}, up to constants, leads to the relation:
\begin{equation}
\lambda_{phys}^{2} >> L_{H}^2 ,
\label{eq:equation3.exxtra}
\end{equation}
where $\lambda_{phys} = \lambda a(t)$.
The inhomogeneous perturbations so, must be at scale greater than the \emph{Hubble Horizon}, hence theoretically they can't actually be observed.

Imposing that our $\delta_{w}$ is now in the range $\left(0,2/3 \right)$, assumption \eqref{eq:equation3.11} is valid, and the system \eqref{eq:equation3.5} reduces to:
\begin{equation}
\begin{cases} 
R^{0}_{0} : \frac{3}{2} k \rho \delta_{w} = 3 \frac{\ddot{a}}{a} + \frac{\ddot{\eta}(t) \theta}{2} + \frac{\dot{a}}{a} \dot{\eta}(t) \theta \\
R^{0}_{\alpha} : - \frac{1}{2} \dot{\eta} (t) \left( \theta_{; \alpha} - \theta^{\gamma}_{\alpha ; \gamma} \right)= k \rho \left(\frac{2}{3} - \delta_{w} \right) u_{\alpha}\\
R^{\alpha}_{\beta} : \frac{1}{2 \sqrt h} \left( \sqrt h K^{\alpha}_{\beta} \right)_{;0} =  k \rho \left[ \left( \frac{2}{3} + \frac{\delta_{w}}{2} \right) \delta^{\alpha}_{\beta} \right] \end{cases}
\label{eq:equation3.12}
\end{equation}
Taking into account the trace of $R^{\alpha}_{\beta}$, we obtain the following equation:
\begin{equation}
R : \frac{1}{2 \sqrt h} \left( \sqrt h K \right)_{,0} =  k \rho \left[ 2 + \frac{3 \delta_{w}}{2} \right] ,
\label{eq:equation3.13}
\end{equation}
the system may be solved by introducing \eqref{eq:equation3.13} in the first equation of \eqref{eq:equation3.12}.
The two equations obtained, respectively for the background and perturbed term, are:
\begin{equation}
\frac{4}{\delta_{w}} \frac{\ddot{a}}{a} = 6 \left( \frac{\dot{a}}{a} \right)^2 ,
\label{eq:equation3.14a}
\end{equation}
\begin{equation}
\ddot{\eta}(t) \theta - \dot{\eta}(t) \theta \frac{\dot{a}}{a} (3 \delta_{w} - 2) = 0 ,
\label{eq:equation3.14b}
\end{equation}
assuming that both $a(t)$ and $\eta(t)$, follow a power law relation in respect of time:
\begin{equation}
a(t) = \left( \frac{t}{t_{0}} \right)^x ,
\label{eq:equation3.15a}
\end{equation}
\begin{equation}
\eta(t) = \left( \frac{t}{t_{0}} \right)^y ,
\label{eq:equation3.15b}
\end{equation}
the first equation \eqref{eq:equation3.14a} reduces to:
\begin{equation}
4x(x -1) = 6 \delta_{w} x^2 \quad \Rightarrow \quad x = 0, \quad x = \frac{1}{1 - \frac{3}{2} \delta_{w}} .
\label{eq:equation3.16a}
\end{equation}
Since the background solution must be an isotropic \emph{FRW} universe, the first solution will be excluded, so that $a(t)$ will assume the same form obtainable with a standard Friedmann approach.

Taking now into account eq. \eqref{eq:equation3.14b}, and considering the solution determined for $a(t)$, the equation for $\eta(t)$ reduces to:
\begin{equation}
\begin{aligned}
&y(y-1) + 2y \frac{1}{1 - \frac{3}{2} \delta_{w}} \left( 1 - \frac{3}{2} \delta_{w} \right) = 0 \\
&\Rightarrow \quad y = 0, \quad y= -1 ,
\label{eq:equation3.16b}
\end{aligned}
\end{equation}
of which, only the second solution may be accepted due to the limit \eqref{eq:equation3.2}.
Considering the solution obtained for $\eta(t)$, the form of the perturbation scale factor $b(t)$ may be determined from the relation
\begin{equation}
\eta(t) = \frac{b^2 (t)}{a^2 (t)} \quad \Rightarrow b^2 (t) = \left( \frac{t}{t_{0}} \right)^{\frac{1 + 3 \delta_{w}/2}{1 - 3 \delta_{w}/ 2}} ,
\label{eq:equation3.17}
\end{equation}
giving the following form, to the metric of the system:
\begin{equation}
h_{\alpha \beta}(t,x) =  \left( \frac{t}{t_{0}} \right)^{\frac{2}{1 - 3 \delta_{w}/2}} \xi_{\alpha \beta} + \left( \frac{t}{t_{0}} \right)^{\frac{1 + 3 \delta_{w}/2}{1 - 3 \delta_{w}/ 2}} \theta_{\alpha \beta} (x) .
\label{eq:equation3.18}
\end{equation}
Furthermore, from the first equation of \eqref{eq:equation3.12}, the energy density reduces to:
\begin{equation}
\begin{aligned}
k \rho = &\frac{2}{\delta_{w} t^2} \frac{1}{1 - \frac{3}{2} \delta_{w}} \left( \frac{1}{1 - \frac{3}{2} \delta_{w}} -1 \right)\\
&- \frac{2}{3 \delta_{w}} \frac{t_{0}}{t^3} \theta \left( \frac{1}{1 - \frac{3}{2} \delta_{w}} -1 \right) ,
\label{eq:equation3.19}
\end{aligned}
\end{equation}
reflecting to the following density contrast:
\begin{equation}
\delta = - \frac{t_{0}}{3 t} \theta \left( 1 - \frac{3}{2} \delta_{w} \right) .
\label{eq:equation3.20}
\end{equation}
The validity of the assumption $u^2 \approx 0$ may be verified from the second equation of \eqref{eq:equation3.12}, leading to the following relation for the tri-velocity:
\begin{equation}
u_{\alpha} = \frac{3}{8} \frac{\delta_{w}}{1 /(1 - 3 \delta_{w}/ 2) - 1} t_{0} ( \theta_{; \alpha} - \theta^{\beta}_{\alpha ; \beta}) ,
\label{eq:equation3.21}
\end{equation}
and consequently:
\begin{equation}
u^2 \propto 1/a^2 (t) = \left( \frac{t}{t_{0}} \right)^{ - \frac{2}{1 - 3 \delta_{w}/2}} \Rightarrow \quad \underset{t \rightarrow \infty}{lim} \quad u^2 = 0 .
\label{eq:equation3.22}
\end{equation}
Using the last equation of \eqref{eq:equation3.12}, it must be observed that the metric perturbation term $\theta^{\alpha}_{\beta}$ must satisfy the condition:
\begin{equation}
\theta^{\alpha}_{\beta} = 1/3 \quad \theta \delta^{\alpha}_{\beta} \quad \Rightarrow \quad \theta^{\alpha}_{\beta ; \alpha} = 1/3 \quad \theta_{; \alpha} \delta^{\alpha}_{\beta} .
\label{eq:equation3.23}
\end{equation}
Lastly, it might be observed that by keeping the tridimensional curvature Ricci's tensor $^3 R^{\alpha}_{\beta}$ in the equations, the system may be solved also for $\delta_{w} = 0$.
The solution obtained, considering a pure temporal point of view, is just a continuous extension of the solution actually derived, as it goes as $a(t) \propto t$ and $\eta(t) \propto 1/t$.

The value $\delta_{w} = -2/3$ instead, will lead to a complete different case, as it represent the case of a pure \emph{Cosmological Constant} universe \cite{Riess:1998cb}.\\

\section{The \emph{Lemaitre-Tolman-Bondi} model for spherically symmetric universes}
\label{sec:4.}

\noindent The \emph{Lemaitre-Tolman-Bondi} model \cite{Peebles:1994xt}, can be thought of as a generalization of the \emph{RW} line element in which the requirement of
homogeneity is dropped, while that of isotropy is kept. De facto the \emph{LTB} model may be isotropic but not homogeneous, due to the fact that by adopting a spherical symmetry, a preferred point is singled out, allowing the space to appear isotropic just by observing the universe from that particular point.
For this reason, it must be said that  the \emph{LTB} approach describes the evolution of a zero-pressure spherical overdensity in the mass distribution, resulting in a spherically symmetric, inhomogeneous solution of the Einstein equations, though the resulting solution is different from the Schwarzschild one because of the non-stationarity.

In the synchronous reference system \eqref{eq:equation2.2}, the spherically symmetric line element for a \emph{LTB} model can be written as:
\begin{equation}
ds^2 = dt^2 - e^{2 \alpha} dr^2 - e^{2 \beta} (d \theta^2 + sin^2 \theta d\phi^2) ,
\label{eq:equation4.1}
\end{equation}
where both $\alpha$ and $\beta$ are functions of time $t$, and the radial spatial coordinate $r$.

The metric \eqref{eq:equation4.1}, will lead to just $3$ independent \emph{Einstein field equations}, that are:
\begin{equation}
\begin{cases} 
kT^{0}_{1} = -2 \dot{\beta}^{\prime} - 2 \dot{\beta} \beta^{\prime}  + 2 \dot{\alpha} \beta^{\prime}\\
kT^{1}_{1} = 2 \ddot{\beta} + 3 \dot{\beta}^2 + e^{-2 \beta} - (\beta^{\prime})^2 e^{-2 \alpha}\\
kT^{0}_{0} = \dot{\beta}^2 + 2 \dot{\alpha} \dot{\beta} + e^{-2 \beta} - e^{-2 \alpha} [2 \beta^{\prime \prime} + 3 (\beta^{\prime})^2 - 2 \alpha^{\prime} \beta^{\prime}]
\end{cases} 
\label{eq:equation4.2}
\end{equation}
where once again, we used the \emph{Einstein constant} $k = 8 \pi G$.
The other non-vanishing equations, will be related to the ones of system \eqref{eq:equation4.2} as:
\begin{equation}
\begin{cases} G^{3}_{3} = G^{2}_{2}\\
G^{2}_{2} = G^{1}_{1} + [G^{1}_{1}]^{\prime}/2 \beta^{\prime} \end{cases}.
\label{eq:equation4.3}
\end{equation}
Originally, this kind of solution was solved under the assumption that the perfect fluid energy-momentum tensor is dominated by pressure-less dust 
$P = 0$ and a cosmological constant term $\Lambda$.
In this scheme, eqs. \eqref{eq:equation4.2} rewrite as:
\begin{equation}
\begin{cases} 0 = -2 \dot{\beta}^{\prime} - 2 \dot{\beta} \beta^{\prime}  + 2 \dot{\alpha} \beta^{\prime}\\
\Lambda = 2 \ddot{\beta} + 3 \dot{\beta}^2 + e^{-2 \beta} - (\beta^{\prime})^2 e^{-2 \alpha}\\
k \rho + \Lambda = \dot{\beta}^2 + 2 \dot{\alpha} \dot{\beta} + e^{-2 \beta} - e^{-2 \alpha} [2 \beta^{\prime \prime} + 3 (\beta^{\prime})^2 - 2 \alpha^{\prime} \beta^{\prime}] \end{cases}
\label{eq:equation4.4}
\end{equation}
where, the $\dot{()}$ and $()^{\prime}$ denote respectively derivatives with respect to time $t$, or radial coordinate $r$.

Since the first equation of \eqref{eq:equation4.4} vanishes, a relation between the two functions $\alpha(r,t)$ and $\beta(r,t)$ may be defined, and it read as:
\begin{equation}
\dot{\beta}^{\prime} /{\beta}^{\prime} = \partial_{t} ln{\beta^{\prime}} = \dot{\alpha} - \dot{\beta} ,
\label{eq:equation4.5}
\end{equation}
which admits the solution:
\begin{equation}
\beta^{\prime} = f(r) e^{\alpha - \beta} ,
\label{eq:equation4.6}
\end{equation}
and consequently:
\begin{equation}
\beta^{\prime} e^{\beta} = \partial_{r} e^{\beta} = f(r) e^{\alpha} .
\label{eq:equation4.7}
\end{equation}
In the last two equations, $f(r)$ was introduced as a generic function of the pure spatial coordinate $r$.
The form of $f(r)$, will be defined upon theoretical considerations about how the inhomogeneity of the model behave in function of $r$, or simply by fitting the resulting model with observational data.

Let us now introduce the commonly used scale factor $a(r, t)$, by adopting the following parametrization:
\begin{equation}
e^{\beta} = r a(r,t), \qquad f(r) = [1 - r^2 K^2]^{1/2} ,
\label{eq:equation4.8}
\end{equation}
where $K = K(r)$, is another free function of the pure spatial coordinate $r$.
It must be observed that although the function $K^2$ has been written as a square, to conform with the standard notation for the isotropic models, $K^2$ can be negative, like in the open isotropic universe.

Using the eqs. \eqref{eq:equation4.7} and \eqref{eq:equation4.8}, the \emph{LTB} line element \eqref{eq:equation4.1} rewrites as follows:
\begin{equation}
ds^2 = dt^2 - \frac{[(ar)^{\prime}]^2}{1 - r^2 K^2} dr^2 - (ar)^2 (d \theta^2 + sin^2 \theta d\phi^2) .
\label{eq:equation4.9}
\end{equation}
It might be observed that, if $a(r,t)$ and $K(r)$ were indipendent of the radial coordinate $r$, the line element \eqref{eq:equation4.9} corresponds 
to the standard \emph{RW} line element.

The remaining field eqs. \eqref{eq:equation4.4}, introducing the relations \eqref{eq:equation4.7} and \eqref{eq:equation4.8}, rewrite now as:
\begin{equation}
(k \rho + \Lambda)[(ar)^3]^{\prime} = 3[\dot{a}^2 ar^3 + ar^3 K^2]^{\prime} ,
\label{eq:equation4.10a}
\end{equation}
\begin{equation}
\Lambda = \frac{2 \ddot{a}}{a} + \frac{\dot{a}^2}{a^2} + \frac{K^2}{a^2} .
\label{eq:equation4.10b}
\end{equation}
Considering in particular eq. \eqref{eq:equation4.10b}, it might be observed that if multiplied by $a^2 \dot{a}$,
it turns into a total time derivative, which can be integrated getting:
\begin{equation}
\dot{a}^2 a + aK^2 - \frac{\Lambda}{3} a^3 = F(r) ,
\label{eq:equation4.11}
\end{equation}
where $F(r)$ is another free function of the pure spatial coordinate $r$, resulting from integration respect to time.\\

\section{The \emph{LTB} approach for spherically symmetric universes in \emph{low-redshift} regime}
\label{sec:5.}

\noindent The \emph{LTB} approach for spherically symmetric universes exposed in the last section, will now be applied in \emph{low-redshift} regime.
By using a \emph{LTB} approach over a \emph{Lifshitz-Khalatnikov Quasi-Isotropic} solution, Einstein equations will appear much easier, as the anisotropy will be limited
to the sole radial coordinate $r$.

While working in \emph{low-redshift} regime for a matter plus cosmological constant universe, Einstein equations remain the same written in \eqref{eq:equation4.2},
moreover from the first of them, the relation between $\alpha(r,t)$ and $\beta(r,t)$ may be determined, following the same procedure of the original model \cite{Peebles:1994xt}.
Using the first equation, the metric will reduce to the form \eqref{eq:equation4.9}, reflecting to the form \eqref{eq:equation4.10a} and \eqref{eq:equation4.10b} for the two remaining Einstein field equations.

To define a small inhomogeneous perturbation, over a flat background Friedmann universe, the scale factor $a(r,t)$ will be defined as:
\begin{equation}
a(r,t) = a_{0}(t) + a_{p}(r,t) ,
\label{eq:equation5.1}
\end{equation}
where $a_{0}(t)$ is the \emph{FRW} unperturbed scale factor, while $a_{p}(r,t)$ represents the inhomogeneous radial perturbation term.
In order to respect the \emph{Cosmological No-Hair conjecture} \cite{Bolejko:2017yqk}, we should later check that the perturbation term satisfies the condition \eqref{eq:equation5.2}.
\begin{equation}
\underset{t \rightarrow \infty}{lim} \frac{a_{p} (r,t)}{a_{0} (t)} = 0 .
\label{eq:equation5.2}
\end{equation}
De facto, the \emph{Cosmological No-Hair conjecture} \cite{Bolejko:2017yqk} states that any Universe with a non-positive spatial curvature and positive cosmological constant (which are indeed verified in our cases), asymptotically evolves towards the De Sitter flat and homogenous universe.
By assuming that the effects of the curvature function $K^2 (r)$, introduced in \eqref{eq:equation4.8}, appear to be negligible in respect to the effects of cosmic accelleration arisen due to the cosmological constant term $\Lambda$, the observable inhomogeneity profile contained in $a_{p}(r,t)$ must asymptotically tend to $0$ for a late stage universe.
Even though the gravitational entropy of our Universe should increase in function of time as a consequence of its long range interaction, such thing may be observed only by looking at our universe in particular critical points where \emph{Large Scale Structures} are situated, hence considering that our $K^2(r)$ only contains average information of the inhomogeneity state, without describing any critical point of our observable universe, the last assumption is valid.

Once the scale factor $a(r,t)$ has been defined following the relation \eqref{eq:equation5.1}, how scale factors $a_{0}(t)$ and $a_{p}(r,t)$ evolves may be obtained from eq. \eqref{eq:equation4.10b}, by separating the $0^{th}$ order terms for the background from the $1^{st}$ order perturbation terms.
We get the following $2$ equations:
\begin{equation}
\frac{2 \ddot{a_{0}}}{a_{0}} + \frac{\dot{a_{0}}^2}{a_{0}^2} = \Lambda ,
\label{eq:equation5.3a}
\end{equation}
\begin{equation}
\frac{2 \ddot{a_{p}}}{a_{0}} - \frac{2 \ddot{a_{0}} a_{p}}{a_{0}^2} + \frac{2 \dot{a_{0}} \dot{a_{p}}}{a_{0}^2} - \frac{2 \dot{a_{0}}^2 a_{p}}{a_{0}^3} + \frac{K^2}{a_{0}^2}  = 0 .
\label{eq:equation5.3b}
\end{equation}
The first equation \eqref{eq:equation5.3a} should give back the standard solution for an \emph{FRW} matter plus cosmological constant universe, we observe that without further assumptions, it admits a class of solutions following the form :
\begin{equation}
a_0(t) = c e^{\sqrt{\frac{\Lambda}{3}} t} \left(d - f e^{-\sqrt{3 \Lambda}t}\right)^{\frac{2}{3}} .
\label{eq:equation5.4}
\end{equation}

In order to define the constants of integrations, we will introduce the new normalized variable $\tilde{t}$ representing the ratio $\tilde{t} = t/t_{0}$.
Assuming that in the following we will denote with $t_{0}$ the value corresponding to the present stage time of the universe, i.e. $t_{0} \approx 1/H_0$, eqt. \eqref{eq:equation5.4} may be rewritten as:
\begin{equation}
a_0(\tilde{t}) = c e^{\sqrt{\frac{\Lambda}{3 H_{0}^{2}}} \tilde{t}} \left(d - f e^{-\sqrt{\frac{3 \Lambda}{H_{0}^{2}}} \tilde{t}}\right)^{\frac{2}{3}} ,
\label{eq:equation5.4bis}
\end{equation}
where by recalling the \emph{critical density} $\rho_{c}$ \cite{Tegmark:2003ud}, we may introduce the \emph{density parameter} $\Omega_{\Lambda} = \frac{\Lambda}{3 H_{0}^{2}}$ for the cosmological constant energy density and we obtain:
\begin{equation}
a_0(\tilde{t}) = c e^{\sqrt{\Omega_{\Lambda}} \tilde{t}} \left(d - f e^{-3\sqrt{ \Omega_{\Lambda}} \tilde{t}}\right)^{\frac{2}{3}} .
\label{eq:equation5.4tris}
\end{equation}
Constants of integration introduced in eqt. \eqref{eq:equation5.4}, may now be defined imposing the two conditions:
\begin{equation}
\underset{\tilde{t} \rightarrow 0}{lim} \quad a(r,\tilde{t}) = 0 ,
\label{eq:equation5.5a}
\end{equation}
\begin{equation}
\underset{\tilde{t} \rightarrow 1}{lim} \quad a(r,\tilde{t}) \approx a_0(\tilde{t}) = 1 ,
\label{eq:equation5.5b}
\end{equation}
reducing eq. \eqref{eq:equation5.4} to:
\begin{equation}
a_0(\tilde{t}) = \frac{e^{\sqrt{\Omega_{\Lambda}} \tilde{t}} \left(1 - e^{-3\sqrt{ \Omega_{\Lambda}} \tilde{t}} \right)^{\frac{2}{3}}}{e^{\sqrt{\Omega_{\Lambda}}} \left(1 - e^{-3\sqrt{ \Omega_{\Lambda}}} \right)^{\frac{2}{3}} } .
\label{eq:equation5.6}
\end{equation}
The two conditions \eqref{eq:equation5.5a} and \eqref{eq:equation5.5b}, represent respectively the assumption that the \emph{cosmologic singularity} \cite{Uzan:2016wji} is approached with a vanishing scale factor, and the standard notation used in astronomy which defines our actual scale factor $a \approx 1$.
To impose the first one, we merely have to set the integration constants of the scale factors in order to have both $a_0(t)$ and $a_p(r,t)$ equals to $0$ for $t = 0$.
The second constraint \eqref{eq:equation5.5b} however is trickier to impose, mainly because as we will see later on in this section, the perturbation term $a_p(r,t)$ doesn't admit any analytical solution. Of course the constraint may still be imposed numerically, however, this would result in a loss of generality of the described model as the structure of $a_0(t)$ and $a_p(r,t)$ would differ in function of the value of $\Omega_{\Lambda}$.
Hence we decided to approximate the total scale factor $a(r,t)$ in \eqref{eq:equation5.5b} with the sole background term $a_0$, this would indeed result in an error proportional to the value of $a_p(r,t)$ in $t = t_0$, yet to describe the theoretical aspects of the model we emphasize that this choice will generalize the obtained solution remarkably.

Moreover, it might be observed that under the precedent assumptions, the background scale factor $a_0$ in the limit:
\begin{equation}
\underset{\Omega_{\Lambda} \rightarrow 0}{lim} \quad a_0(\tilde{t}) \approx \tilde{t}^{2/3},
\label{eq:equation5.7}
\end{equation}
shows that for low values of $\Omega_{\Lambda}$, the model behaves as a pure matter dominated universe, as we would normally expect from the definition of our model.

Now taking into account eq. \eqref{eq:equation5.3b}, suggests a factorization as follows:
\begin{equation}
a_{p} (r,\tilde{t}) = b(\tilde{t}) K^{2} (r) ,
\label{eq:equation5.8}
\end{equation}
where $b(\tilde{t})$ represent the temporal dependance of the perturbation scale factor $a(r,\tilde{t})$ and $K^2(r)$ may be negative according to the standard definition of the \emph{LTB} model \eqref{eq:equation4.8}.
It might be observed that a linearized solution of the perturbed equation \eqref{eq:equation5.3b} in terms of a factorization in a function of the radial coordiate, 
times a function depending on time is, in principle, not a general one representing the available Universe inhomogeneity.
However, we stress that a careful analysis of the equation structure suggests such a separation of variables as a natural choice, mainly due to the linearization 
procedure of the dynamics.
Furthermore, we are also lead to such a natural choice because of the necessity to compare the obtained 
result in spherical symmetry with the quasi-isotropic solution of section \ref{sec:3.}, which is, ab initio, 
postulated via a factorization of the space and time dependence of the perturbation.

The factorization \eqref{eq:equation5.8}, leads to the following equation:
\begin{equation}
\begin{aligned}
&\ddot{b}(\tilde{t}) + \dot{b}(\tilde{t}) \left( \sqrt{\Omega_{\Lambda}} + \frac{2 \sqrt{\Omega_{\Lambda}} e^{- 3\sqrt{\Omega_{\Lambda}} \tilde{t}}}{1 - e^{- 3\sqrt{\Omega_{\Lambda}} \tilde{t}}}\right) \\
&- 2b(\tilde{t}) \left( \Omega_{\Lambda} + \frac{ \Omega_{\Lambda} e^{-3 \sqrt{ \Omega_{\Lambda}} \tilde{t}}}{1 - e^{-3 \sqrt{ \Omega_{\Lambda}} \tilde{t}}} + \frac{ \Omega_{\Lambda} e^{-6 \sqrt{ \Omega_{\Lambda}} \tilde{t}}}{ \left(1 - e^{-3 \sqrt{\Omega_{\Lambda}} \tilde{t}} \right)^{2}} \right)\\
&+ \frac{e^{\sqrt{\Omega_{\Lambda}}} \left(1 - e^{-3 \sqrt{\Omega_{\Lambda}}} \right)^{\frac{2}{3}}}{2 e^{\sqrt{\Omega_{\Lambda}} \tilde{t}} \left(1 - e^{-3 \sqrt{\Omega_{\Lambda}}\tilde{t}}\right)^{\frac{2}{3}}} = 0 ,
\label{eq:equation5.9}
\end{aligned}
\end{equation}
that doesn't admit any analytic solution.

In order to solve that a \href{https://github.com/KuZa91/LTB-solution-for-perturbation-scale-factor-equation/blob/master/LTBMatterCostant.f90}{software} was built in \emph{Fortran} $90$ that integrate equation \eqref{eq:equation5.9} using an \emph{Explicit Euler} method.

Being a second order differential equation, we need two starting conditions in order to run simulations. 
We decided to set up $b(0) = 0$ and $\dot{b}(0) = \dot{a_0}(0)$.
The first condition comes naturally by imposing that the \emph{cosmologic singularity} \cite{Uzan:2016wji} is approached with a vanishing scale factor, 
for what concerns the second one, we decided to couple the evolution of the perturbation scale factor $b$ to the background scale factor $a_0$ in the first stage of the evolution of the universe, we may hence state that in the limit $\tilde{t} \approx 0$ the model behaves as follow:
\begin{equation}
\underset{\tilde{t} \rightarrow 0}{lim} \quad a(r,\tilde{t}) = a_{0}(\tilde{t}) \left(1 + K^{2} (r) \right) .
\label{eq:equation5.1BIS}
\end{equation}
This assumption was also motivated by assuming a smooth evolution of the $\eta(\tilde{t}) = b^2 (\tilde{t}) / a_0^2 (\tilde{t})$ variable around $\tilde{t}= 0$.

\begin{figure}[h]
\centering 
\includegraphics[width=\columnwidth]{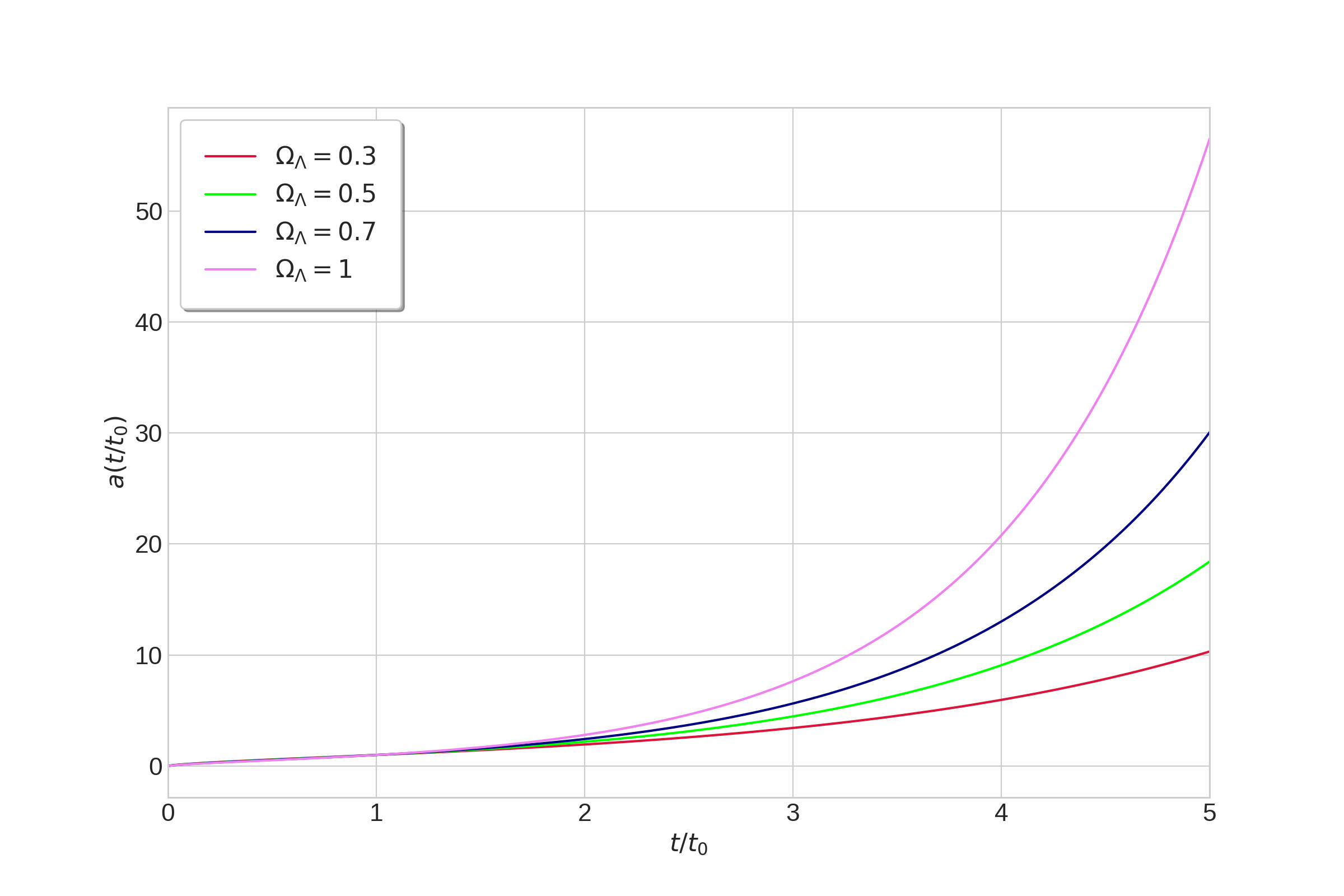} 
\caption{\it Evolution of $a(t)$ scale factor in function of $\Omega_{\Lambda}$ for a \emph{LTB} model.\\}
\label{fig:figura5.1}
\end{figure}

\begin{figure}[h]
\centering 
\includegraphics[width=\columnwidth]{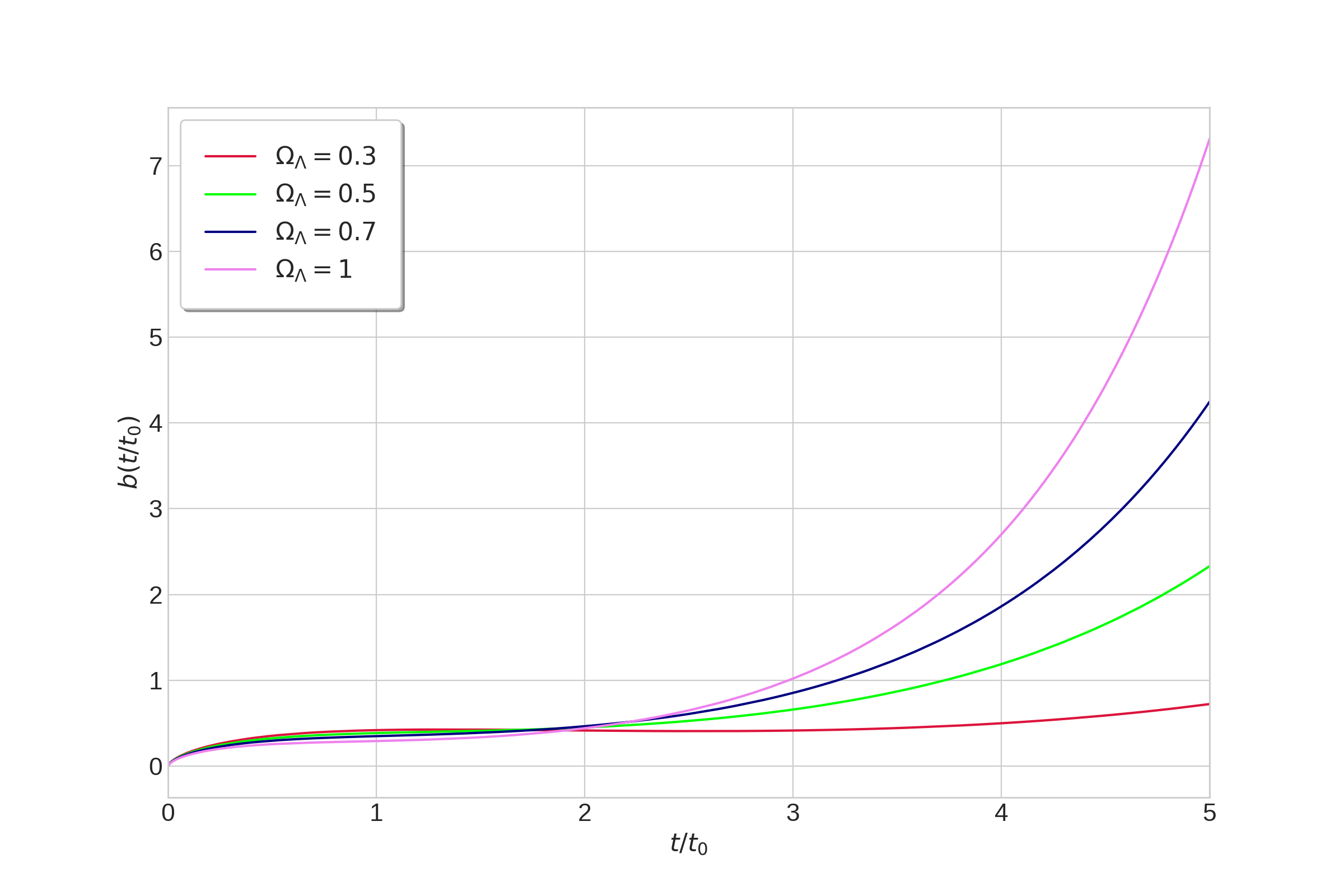} 
\caption{\it Evolution of $b(t)$ scale factor in function of $\Omega_{\Lambda}$ for a \emph{LTB} model.\\}
\label{fig:figura5.2}
\end{figure}

\begin{figure}[h]
\centering 
\includegraphics[width=\columnwidth]{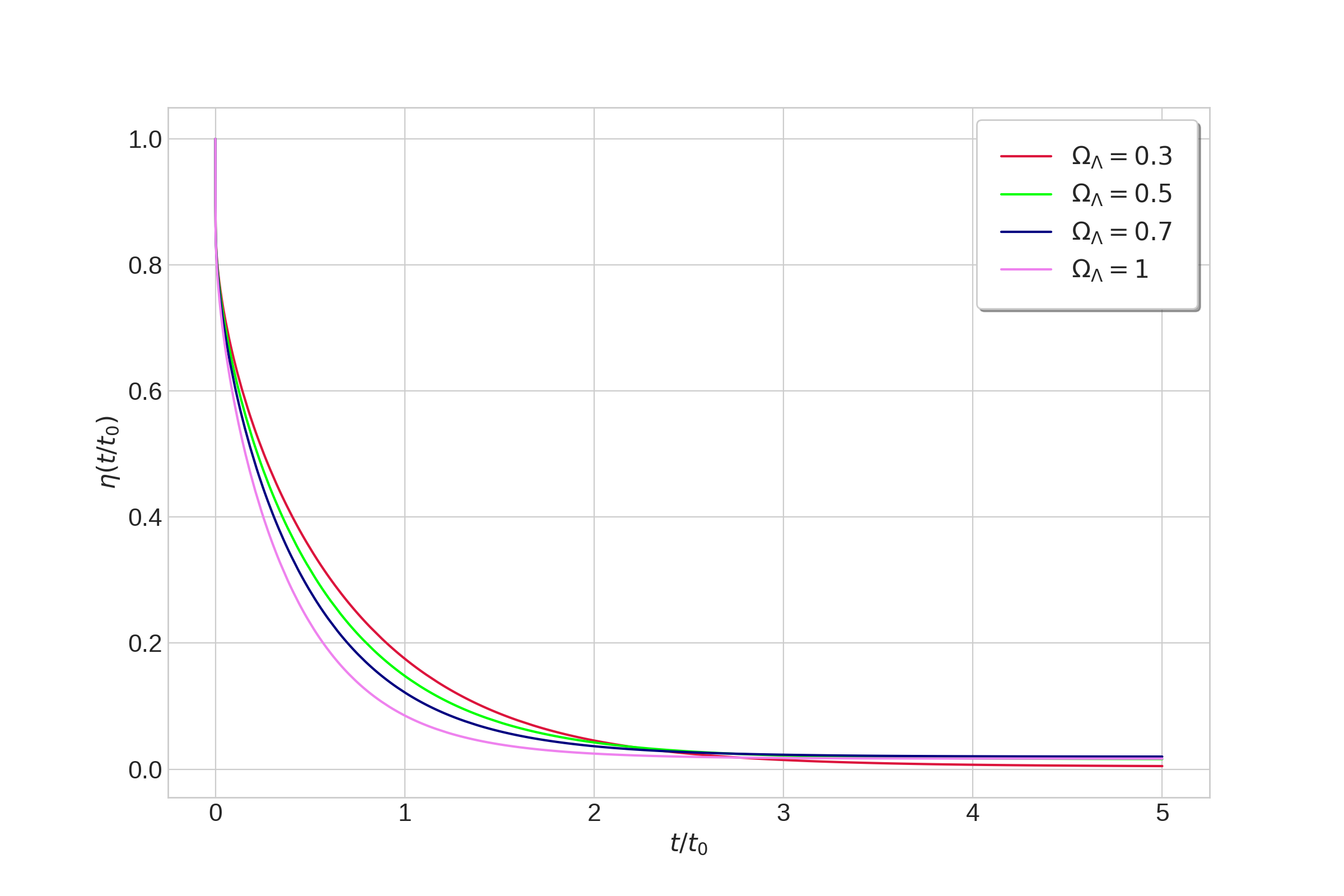} 
\caption{\it Evolution of $\eta(t)$ scale factors ratio in function of $\Omega_{\Lambda}$ for a \emph{LTB} model.\\}
\label{fig:figura5.3}
\end{figure}

Equation \eqref{eq:equation5.9} was solved using $4$ different values of $\Omega_{\Lambda}$, in particular, the values tested were $\Omega_{\Lambda} = 0.3, 0.5,0.7,1$ and the results obtained from the simulation for $a_0(\tilde{t})$ and $b(\tilde{t})$ in function of $\Omega_{\Lambda}$ will be shown in figures \ref{fig:figura5.1} and \ref{fig:figura5.2}.
In particular, we might observe from fig. \ref{fig:figura5.2} that the evolution over time of inhomogeneities, given by the evolution of the perturbation scale factor $b(\tilde{t})$ is rather small compared to the homogeneous background given by $a_0(\tilde{t})$.
This becomes more clear by observing the results obtained from the simulation for $\eta(\tilde{t}) = b^2 (\tilde{t}) / a_0^2 (\tilde{t})$, which behaves as reported in figure \ref{fig:figura5.3}.
The last graph shows that the condition of eq. \eqref{eq:equation5.2} is valid, in fact, for all the values of $\Omega_{\Lambda}$ considering $\eta(\tilde{t})$ becomes smaller than $0.2$ already at $1 \; t/t_{0}$, reflecting to a ratio $a_{p}/a_{0} \approx 0.4$ in correspondance to the same value.
The value of $\eta(\tilde{t})$, will then asymptotically converge to $0$ in the limit $\tilde{t} \rightarrow \infty$, in agreement with the \emph{Cosmological No-Hair conjecture} \cite{Bolejko:2017yqk}.

To further analyze the properties of the perturbation scale factor $b(\tilde{t})$, we decided to check if a minimum value $\Omega_{\Lambda min}$ exists in order to have non-vanishing, and hence non-negative, values of the latter in the considered time interval.
The cosmological implications of a vanishing or negative perturbation scale factor will not be discussed in detail in this paper, as it will occur for values of $\Omega_{\Lambda}$ that will be 
out of the scope when trying to describe the present-stage fiducial observations of our universe.   
De facto, by looking at fig. \ref{fig:figura5.2} we see that, if such a value exist, it should be in the range $\Omega_{\Lambda min} \in \left(0.,0.3 \right)$, as we already tested values of $\Omega_{\Lambda} \geq 0.3$.
In order to give a better estimation of the real value, 
we built \href{https://github.com/KuZa91/LTB-solution-for-perturbation-scale-factor-equation/blob/master/MinLambdaEstimate.f90}{another software} in \emph{Fortran} $90$ that adopts a 
\emph{Divide et Impera} method to converge to said value. At each step, the software splits the fiducial interval in two equal size ranges, and chooses one of the two 
as the new fiducial interval by looking at the evolution of the perturbation scale factor at the boundary point between them.

\begin{figure}[h]
\centering 
\includegraphics[width=\columnwidth]{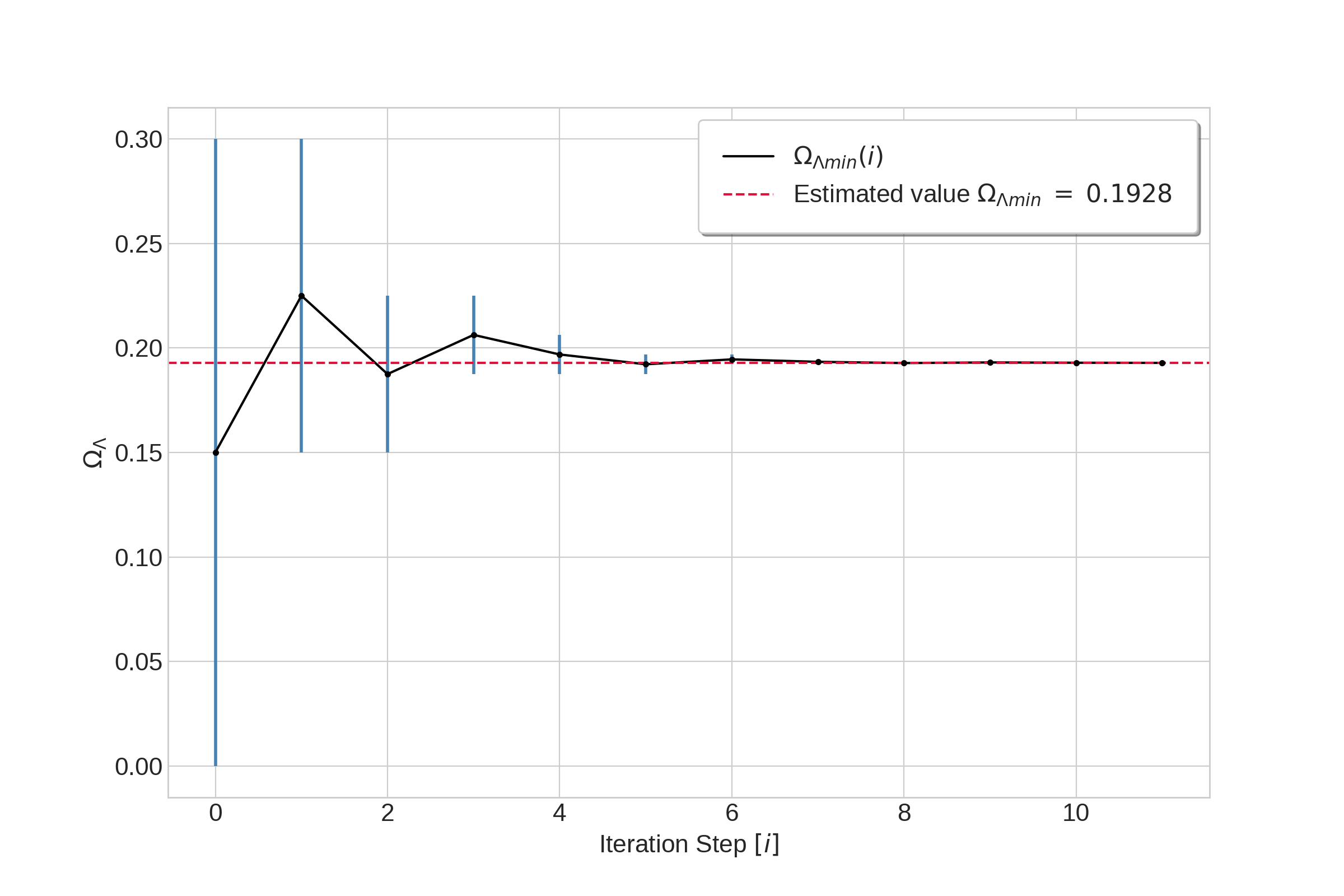} 
\caption{\it Estimation of the value of $\Omega_{\Lambda min}$ in order to have non-vanishing inhomogeneous perturbations in the time interval $\tilde{t} \in \left[0.,5. \right]$. We see that on each step, the interval containing $\Lambda_{min}$ gets split in half resulting in a convergence precision in function of the step $i$ evolving as $L_0 / 2^i$, where $L_0$ is the length of the initial range.\\}
\label{fig:figura5.4}
\end{figure}

We found that inhomogeneous perturbations of the model will vanish in a time smaller than $\tilde{t} = 5.$ for models having a value of $\Omega_{\Lambda}$ smaller than $\Omega_{\Lambda min} = 0.1928 \pm 1*10^{-4}$, the convergence of the \href{https://github.com/KuZa91/LTB-solution-for-perturbation-scale-factor-equation/blob/master/MinLambdaEstimate.f90}{software} to the previously stated value is shown in fig. \ref{fig:figura5.4}.

To conclude the section, eq. \eqref{eq:equation4.10a}, under assumption \eqref{eq:equation5.1}, \eqref{eq:equation5.2} and \eqref{eq:equation5.8}, reduces to the following form in function of $K(r)$:
\begin{equation}
\begin{aligned}
k\rho + \Lambda \; = \; &  3\left( \frac{\dot{a}_{0}}{a_{0}} \right)^{2} - 6 \frac{\dot{a}_{0}^{2}}{a_{0}^{3}}bK^{2}(r) -4r \frac{\dot{a}_{0}^{2}}{a_{0}^{3}}b K(r) K^{\prime}(r)\\
&+ 6\frac{\dot{a}_{0}}{a_{0}^{2}} \dot{b} K^{2}(r) + 4r \frac{\dot{a}_{0}}{a_{0}^{2}} \dot{b} K(r) K^{\prime}(r) + 3 \frac{K^{2}(r)}{a_{0}^{2}} \\
& - 6b \frac{K^{4}(r)}{a_{0}^{3}} -8rb\frac{K^{3}(r) K^{\prime}(r)}{a_{0}^{3}} + 2 r \frac{K(r) K^{\prime}(r)}{a_{0}^{2}} \\
&-4 r^{2} b\frac{K^{2}(r) (K^{\prime}(r))^{2}}{a_{0}^{3}} ,
\label{eq:equation5.10}
\end{aligned}
\end{equation}
where by assuming that $K(r)$ has a profile as follows:
\begin{equation}
K(r) \sim \frac{K_{0}}{(1 + r)^\gamma} ,
\label{eq:equation5.11}
\end{equation}
in which both $K_{0}$ and $\gamma$ are parameters that need to be determined from observational data,
higher orders term in $K(r)$ of eq. \eqref{eq:equation5.10} may be neglected, leading to the much simpler form for the density profile in function of the radial coordinate $r$:
\begin{equation}
\begin{aligned}
k\rho + \Lambda \sim & 3\left( \frac{\dot{a}_{0}}{a_{0}} \right)^{2} + \frac{K_{0}^{2}}{(1 + r)^{2 \gamma}}\left[ 6\frac{\dot{a}_{0}}{a_{0}^{2}} \dot{b} -6 \frac{\dot{a}_{0}^{2}}{a_{0}^{3}}b + \frac{3}{a_{0}^{2}} \right] \\
&+\frac{\gamma r K_{0}^{2}}{(1 + r)^{2 \gamma + 1}} \left[4 \frac{\dot{a}_{0}^{2}}{a_{0}^{3}}b - 4 \frac{\dot{a}_{0}}{a_{0}^{2}} \dot{b} - \frac{2}{a_{0}^{2}} \right] .
\label{eq:equation5.12}
\end{aligned}
\end{equation}
The determination of the best form for the function $K(r)$, in order to represent the inhomogeneities of the universe we live in, will not be a subject of study in this paper and consequently, the exact form for the metric \eqref{eq:equation4.9} and for the energy density \eqref{eq:equation5.10} will not be defined here.
However, by observing inhomogeneities of our local universe, the form of $K(r)$ may be easily defined upon opportune fits with the experimental data, hence the best solution for this model in order to represent our present-stage universe may be tested by incoming missions like \emph{Euclid}\cite{Amendola:2016saw}.
\subsection{Comparison of the \emph{LTB} analysis with the actual universe}
\label{ssec:1.}
\noindent The \emph{LTB} model for matter and cosmological constant universes will now be compared with observational data to fit the model that best describe our actual universe, even though in the following, the function $K(r)$ will remain a free function of spatial coordinate $r$. Although the matching will be roughly obtained in the flat \emph{FRW} limit, a more accurate comparison will not be subject of this paper as a complete match requires observational data of the inhomogeneities of our universe, to define the best form for $K(r)$. A detailed discussion on how to rigorously fit an \emph{LTB} model to observational data, may be found in external papers such as \cite{Redlich:2014gga},\cite{McClure:2007hy} and \cite{Celerier:2012sj}.

By now, observations suggest that our universe is composed of $\approx 30\%$ matter, and for the remaining $\approx 70\%$, by \emph{D.E}/\emph{cosmological constant} \cite{Tegmark:2003ud}.
At first, the model in the limit $t \rightarrow t_{0}$, where $t_{0}$ represents present stage time, must return the same density composition. To impose such a thing, eq. \eqref{eq:equation5.10} will be taken into account, in particular, it might be observed from figure \ref{fig:figura5.1} and \ref{fig:figura5.2} that the perturbation scale factor $b(t)$ appears to be in first approximation negligible in respect to the background scale factor $a_{0}(t)$. Consequently, we may approximate eq. \eqref{eq:equation5.1} to the form :
\begin{equation}
a(r,t) \approx a_{0}(t) \quad \Rightarrow \quad b(t) \approx 0 .
\label{eq:equation6.1}
\end{equation}
Furthermore, by imposing that the universe is averagely flat, which from eq. \eqref{eq:equation4.9} turns to $r^2 K^{2} (r) \approx 0$ i.e. $K^{2} (r) \approx 0$, eq. \eqref{eq:equation4.10a} reduce to the standard form for the energy density obtainable from a \emph{FRW} Universe:
\begin{equation}
k \rho = \frac{k \rho_{m}}{a^3 (t)} + \Lambda ,
\label{eq:equation6.2}
\end{equation}
where $\rho$ represents our universe total energy density, $\rho_{m}$ is the fraction of total density as matter, and $\Lambda$ is the fraction as cosmological constant.

Switching to normalized densities by using \emph{Critical density} $\rho_{c}$ \cite{Tegmark:2003ud}, the actual total energy density gets the value $\Omega_{tot}(t_{0})= \rho(t_{0})/\rho_c = 1$.

Moreover taking into account the condition imposed on our $a_0(t)$ \eqref{eq:equation5.5b}, the values of $\Omega_m = \rho_{m}/ \rho_c$ and $\Omega_{\Lambda} = \Lambda / \rho_c$ may be easily defined using \eqref{eq:equation6.2}, getting:
\begin{equation}
1 \quad = \quad \Omega_{tot}(t_{0}) \quad = \quad \frac{\Omega_m}{a^3 (t_{0})} + \Omega_{\Lambda} ,
\label{eq:equation6.3}
\end{equation}
which, recalling that we have $a(t_0) = 1$ and the proportions between matter and cosmological costant should respect the ones actually observed \cite{Tegmark:2003ud}, implies $\Omega_{m} = 0.3$ and $\Omega_{\Lambda} = 0.7$.

The behavior of the background and perturbation scale factor for an \emph{LTB} universe model having $\Omega_{\Lambda} = 0.7$, as well as the behavior of the scale factors squared ratio $\eta(t)$, are already shown respectively in figures \ref{fig:figura5.1}, \ref{fig:figura5.2} and \ref{fig:figura5.3}. 
However, as it is impossible to define an analytical form for $b(t)$ and $\eta(t)$, an approximated shape for their analytical form will be estimated from the simulated data by using \href{https://www.originlab.com/}{Origin} software to fit the numerical data.

\begin{figure}[h]
\centering 
\includegraphics[width=\columnwidth]{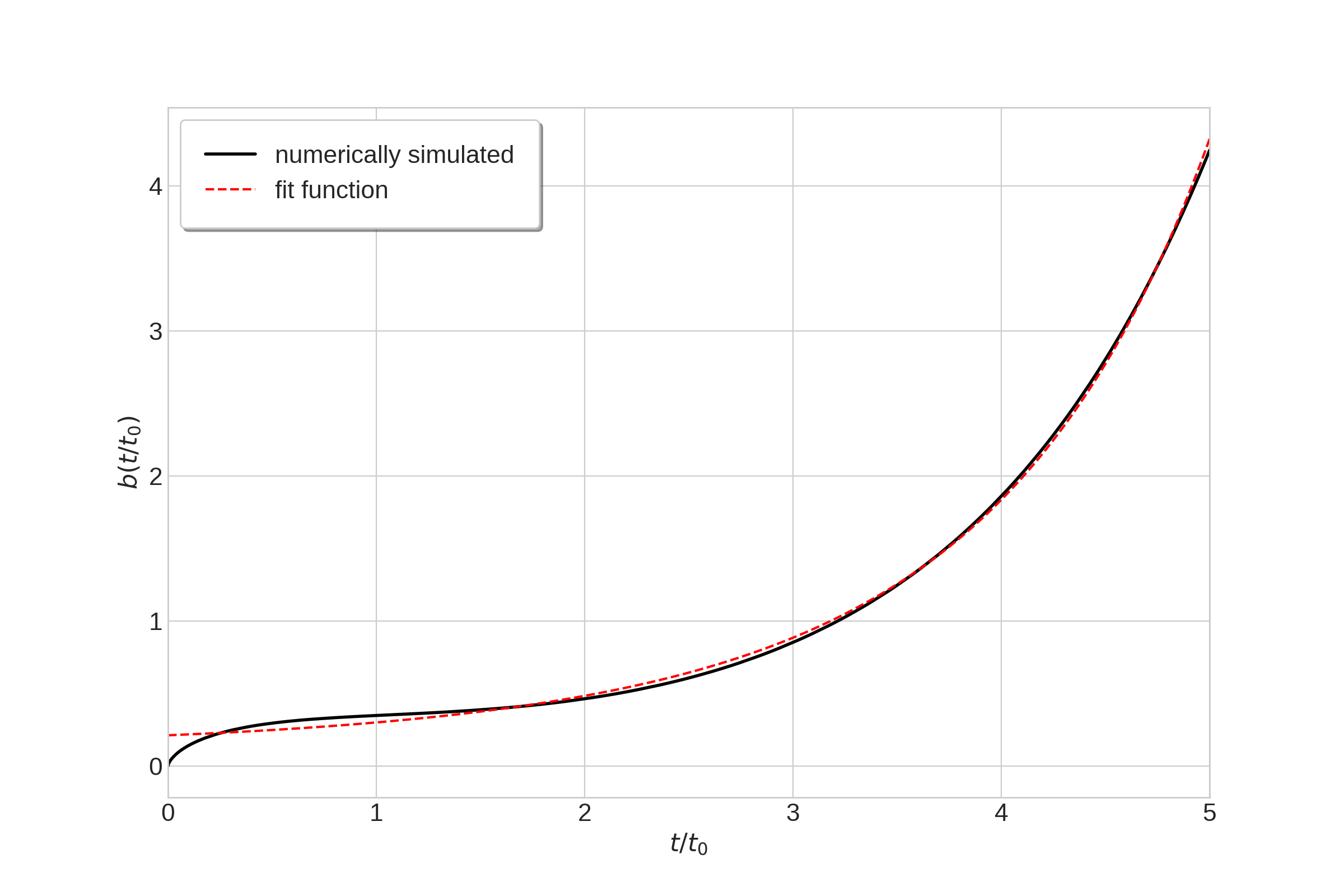} 
\caption{\it Comparison between the numerical $b(t)$ (in black) and the exponential fit function (in dashed red), for a universe having $\Omega_{\Lambda} = 0.7$.\\}
\label{fig:figura6.1}
\end{figure}

For what concerns $b(\tilde{t})$, the best fit found is with a function having the shape described in \eqref{eq:equation6.4}:
\begin{equation}
b(\tilde{t}) = e^{a + b\tilde{t}+ c\tilde{t}^2} ,
\label{eq:equation6.4}
\end{equation}
where the estimated coefficients are shown in tab. \ref{tab:tabular1}.
\begin{table}
\centering
\begin{tabular}{cc}
Coefficients & Value \\ \hline
\\
$a$ & $ -1.550 \; \pm \; 0.001$\\
\\
$b$ & $ 0.2856 \; \pm \; 8.18*10^{-4}$\\
\\
$c$ & $ 6.34*10^{-2} \; \pm \; 1.14*10^{-4}$\\
\\ \hline
\end{tabular}
\caption{Table of the coefficients for $b(t)$ exponential fit function.}
\label{tab:tabular1}
\end{table}
In figure \ref{fig:figura6.1}, we show a comparison between the numerical data of the perturbation scale factor $b(t)$ and the fit function.

\begin{figure}[h]
\centering 
\includegraphics[width=\columnwidth]{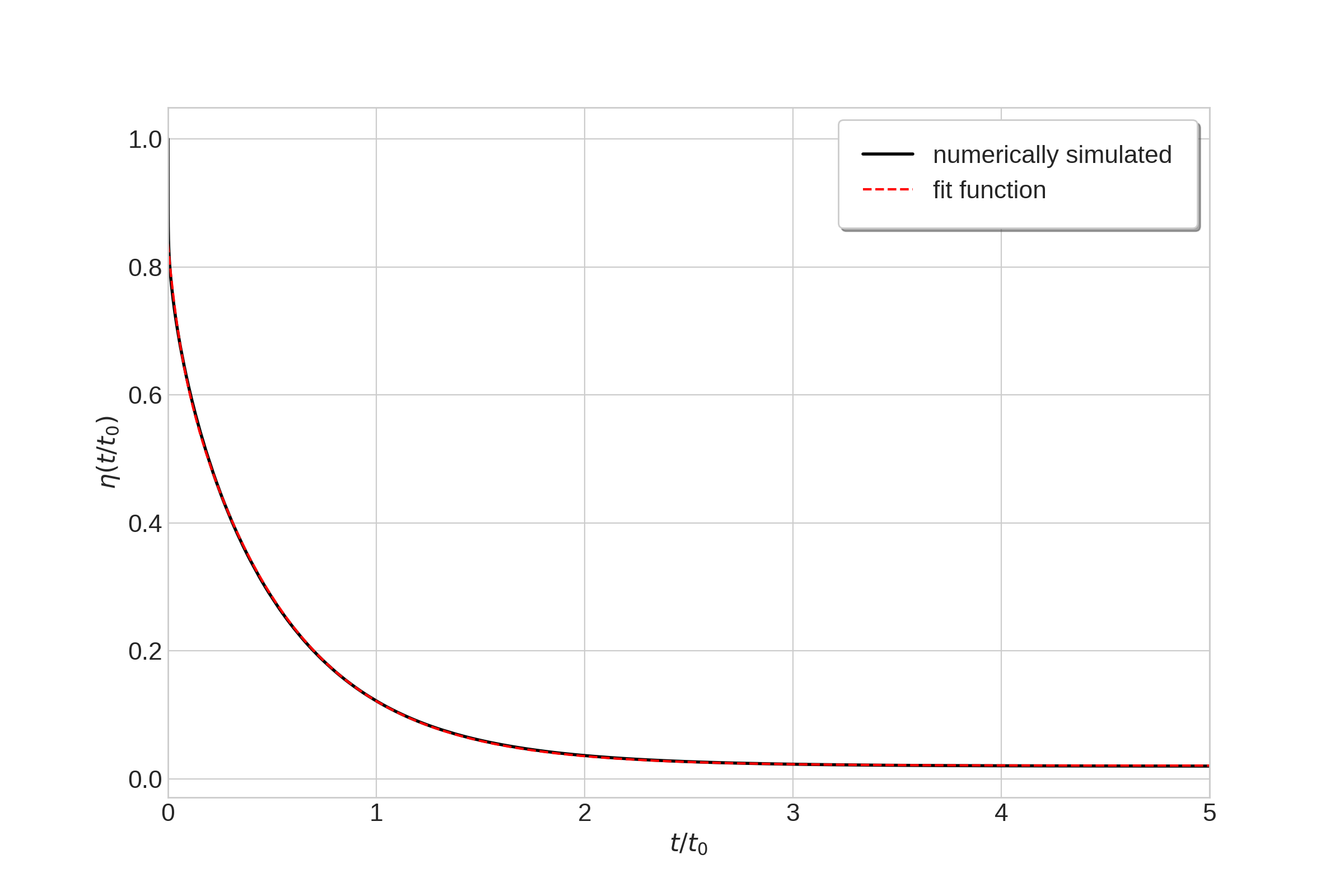} 
\caption{\it Comparison between the numerical $\eta(t)$ (in black) and the exponential fit function (in dashed red), for a universe having $\Omega_{\Lambda} = 0.7$.\\}
\label{fig:figura6.2}
\end{figure}

We may now proceed in the same way for $\eta(\tilde{t})$. Let us start by defining the analytical function used to fit the data in \eqref{eq:equation6.5}.
\begin{equation}
\eta(\tilde{t}) = \eta_0 + A_1 \left(1 -e^{-\tilde{t}/\tau_1} \right) + A_2 \left(1 - e^{-\tilde{t}/\tau_2} \right) .
\label{eq:equation6.5}
\end{equation}
The values of the coefficients, for the best fit, were set to the values reported in tab. \ref{tab:tabular2},
\begin{table}
\centering
\begin{tabular}{cc}
Coefficients & Value \\ \hline
\\
$\eta_0$ & $ 0.8360 \; \pm \; 1.68*10^{-4}$\\
\\
$A_1$ & $ -0.1266 \; \pm \; 1.75*10^{-4}$\\
\\
$A_2$ & $ -0.6887 \; \pm \; 1.16*10^{-4}$\\
\\
$\tau_1$ & $ 0.0530 \; \pm \; 1.45*10^{-4}$\\
\\
$\tau_2$ & $ 0.5206 \; \pm \; 1.00*10^{-4}$\\
\\ \hline
\end{tabular}
\caption{Table of the coefficients for $\eta(t)$ exponential fit function.}
\label{tab:tabular2}
\end{table}
and as before, a comparison between the fit function determined with \href{https://www.originlab.com/}{Origin}, and the numerical simulated data for the $\eta(t)$ function, will be shown in figure \ref{fig:figura6.2}.\\
\subsection{Effects of Inhomegeneities on the measuration of Hubble Constant $H_{0}$}
\label{ssec:2.}

\noindent In this last subsection, in order to elucidate the idea that the small inhomogeneities of the obtained perturbed sperically symmetric model
are able to account for the discrepancies in the detected value of the \emph{Hubble constant} $H_0$ \cite{Hubble:1929ig} by different 
observational tasks \cite{Chen:2017rfc} \cite{Wong:2019kwg}, we now develop some simple qualitative considerations, 
in principle to be restated in an exact calculation of the photon propagation.

From the spherically symmetric line element \eqref{eq:equation4.9}, 
it is easy to realize that, if we split the scale factor as shown in \eqref{eq:equation5.1} where the perturbation term has the form
defined in \eqref{eq:equation5.8}, the line element reduces to:
\begin{equation}
\begin{aligned}
ds^2 =& dt^2 - \frac{[a_{0}(t) + b(t)K^{2}(r) + r b(t) K(r) K^{\prime}(r)]^2}{1 - r^2 K^2} dr^2\\
& -(a_{0}(t) + b(t) K^{2} (r))^{2} r^{2}(d \theta^2 + sin^2 \theta d\phi^2) .
\label{eq:equation7.1}
\end{aligned}
\end{equation}
Assuming then that the condition $|r K(r) K^{\prime}(r)| \ll |K^{2}(r)|$ holds,
eq. \eqref{eq:equation7.1} may be written as follows:
{ 
\begin{equation}
\begin{aligned}
ds^2 = & dt^2 - (a_{0}(t) + b(t)K^{2}(r) )^2 \\
&\left[ \frac{1}{1 - r^2 K^{2}(r)} dr^2 -  r^{2}(d \theta^2 + sin^2 \theta d\phi^2) \right] ,
\label{eq:equation7.2}
\end{aligned}
\end{equation}}
where we factorized the space line element, in terms of the scale factor previously defined in \eqref{eq:equation5.1} times a pure radial-depending 3-metric. 
Under this assumptions, we can extract from the space depending part of the $3$-metric 
(de facto the one of the isotropic Universe apart from the small dependence of $K^2$ on $r$), a common scale factor which may be compared with the ones 
defined for the \emph{FRW} universes.

If we now define a \emph{pseudo-expansion rate} of the Universe as: 
\begin{equation}
\begin{aligned}
\tilde{H}(t,r) =& \frac{\partial _{t} a}{a} = \frac{\dot{a}_{0} + \dot{b} K^{2}}{a_0 + b K^2}\\ 
&\approx \frac{\dot{a}_{0}}{a_{0}} + \left[ \frac{\dot{b}}{a_{0}} - \frac{\dot{a}_{0} b}{a_{0}^{2}} \right] K^{2} = H_{0} + \delta H (t,r) ,
\label{eq:equation7.3}
\end{aligned}
\end{equation}
we observe that for a weakly inhomogeneous Universe, which may indeed be the case of our present stage observable Universe for spatial scales smaller than $60 Mpc/h$, 
this scheme for the expansion rate holds for a given (sufficiently small) radius $r$ around the observer in the 
center of symmetry.
Thus, it becomes clear that a Hubble law holds only for photons emitted from the same spherical surface, i.e. at the same distance from the centre of symmetry.

By assuming that the homogeneous Universe is a reliable model, we are disregarding this dependence of the expansion 
rate on the radial coordinate and we are, de facto using an average value of the Hubble constant. 
In our model the inhomogenity effect is thought small, say of the order of $10$ percent, thus local 
measurements of $H_0$ could be affected by this order of approximation on the homogeneity hypothesis, 
therefore deviating from the measurements of $H_0$ from the \emph{CMB}, 
based on very large scale properties of the actual Universe \cite{Ade:2015rim}.
For a more detailed discussion on how the inhomogeneities of our universe may affect Hubble constant measurements, see \cite{Odderskov:2016rro} \cite{Odderskov:2017ivg} \cite{Enqvist:2007vb}.\\

\section{Concluding Remarks}
\label{sec:8.}
\noindent We analyze two different, but complementary algorithms to deal with small inhomogeneus corrections to the isotropic Universe: 
on one hand, we studied the so-called \emph{quasi-isotropic solution}, as implemented to a late dynamics, 
on the other, we study a \emph{Lemaitre-Tolamann-Bondi} spherically symmetric solution, 
containing only small deviations depending on the pure radial coordinate.
We considered in both cases sources in the form of a perfect fluid, but 
while for the quasi-isotropic case we consider a dark energy equation of state with $-1< w<-1/3$, 
the spherically symmetric solution contains two different contributions, a matter fluid and a cosmological constant, respectively.

The basic result of our analysis, is demonstrating that the presence of a real dark energy contribution, 
prevent the possibility to deal with physical scales of the inhomogeneous correction being smaller than the actual Hubble scale of the Universe.
This constraint, comes from the necessity to rule out of the solution the spatial curvature contribution (due to inhomogeneous corrections), 
and it has very deep implications: the obtained perturbed solution is characterized by perturbations evolving only from a kinematical point of view, 
but unaffected by microphysical processes and, de facto they are not observable at the present time.
Clearly, this result can not be considered as a general one for two basic reasons :
\begin{itemize}
	\item It depends on the details of the solution construction, i.e. by retaining curvature effects of the perturbations, more general regimes may exist;

	\item  The perturbation component of a quasi-isotropic solution is naturally factorized into the product of a space dependent and a time dependent function respectively, and this influences the obtained dynamics.\\   
\end{itemize}
For a discussion of related problems regarding sub-Hubble inhomogeneities see \cite{Wiseman:2010kp}.

The situation is different for the Lemaitre-Tolman-Bondi model, where 
we actually consider spherically symmetric deviations to the background dynamics, underlying the $\Lambda CDM$ model for the actual Universe.
We construct a inhomogeneous perturbation to the isotropic cosmology whose spatial dependence, i.e. spectrum, 
is not fixed by the solution method, retaining a useful degree of freedom for fitting different physical situations.

Apart from the conceptual difference qualitatively emerging in the present study between the two used algorithms,
the main merit of this work is outlining that in the \emph{LTB} case, the inhomogeneous perturbation results in  a consistent solution with late time sub-Hubble inhomogeneities.
In fact, the possibility to check the geodesic dynamics on different weakly inhomogeneous Universe models 
offers an interesting tool to test some physical properties of the actual low-redshift Universe.

In particular, as briefly shown in subsection \ref{sec:8.}, we suggest that the Lemaitre-Tolman-Bondi model studied above could be adopted to try to account, by using weak inhomogeneity profiles, the discrepancy existing between the value of the \emph{Hubble constant} $H_0$ \cite{Hubble:1929ig} 
as it is measured by \emph{WMAP}, \emph{Planck} Satellities and by the ground based surveys \cite{Chen:2017rfc} \cite{Wong:2019kwg} \cite{Lasenby:1998zx}.\\
The elimination of such a discrepancy by the proposed scenario, could put a limit on the local inhomogeneity profile of the actual Universe, 
possibly tested by the incoming mission \emph{Euclid}\cite{Amendola:2016saw}. \\
\begin{acknowledgments}
\noindent We thank \emph{Germano Nardini} and \emph{Alex B. Nielsen} for the help in writing and fixing some theoretical aspects of the third version of this paper.
Furthermore we thank \emph{Tiziano Schiavone}, for reporting an error in eqt. \eqref{eq:equation5.9}, and \emph{Kristine Maeland} for the last revision of the text.
All the figures were realized by using \href{https://www.anaconda.com/}{Anaconda Distribution} together with the \href{https://seaborn.pydata.org/index.html}{Seaborn visualization library}, additional information on how to use the previously described environment may be found in the book \cite{10.5555/3133359}.

\end{acknowledgments}

\bibliography{Zib.bib}

\end{document}